\documentstyle[12pt]{article}
\newcommand{\bea}{\begin{eqnarray}}
\newcommand{\beq}{\begin{equation}}
\newcommand{\eea}{\end{eqnarray}}
\newcommand{\eeq}{\end{equation}}

\def\ga{\mathrel{\mathpalette\fun >}}
\def\fun#1#2{\lower3.6pt\vbox{\baselineskip0pt\lineskip.9pt
  \ialign{$\mathsurround=0pt#1\hfil##\hfil$\crcr#2\crcr\sim\crcr}}}
\begin{document}
\pagestyle{empty}
\begin{flushright} UPR-541T \end{flushright}
\begin{center}

\vspace{.5in}
{\bf CHALLENGES FOR SUPERSTRING COSMOLOGY  }
\\

\vspace{.3in}
 {\sc Ram Brustein and Paul J. Steinhardt }\\
{ Department of Physics, \\
University of Pennsylvania, Philadelphia, PA  19104}\\
%\today \\
\vspace{1.2in}
%\centerline{\psfig{file=UPLOGO.epsf.ps,height=1.5in,width=1.5in}}
%\epsfscale=700 \centerline{\ \ \ \epsfbox{uplogocopy.epsf}}
%

\vspace{.5in}
\noindent
{\bf ABSTRACT}
\end{center}
We consider whether current notions about  superstring theory below the
Planck scale are compatible with cosmology.  We find that the anticipated
form for the dilaton interaction creates a serious roadblock for inflation
and makes it unlikely that the universe ever reaches a state with
zero cosmological constant and time-independent  gravitational constant.

\vspace{.1in}
\noindent

\newpage
\pagestyle{plain}
\setcounter{page}{1}

 Superstring theories have been extensively studied as models of unified
theories
\cite{gsw}.
Effective field theories  in four dimensions are used
to describe string theory  below the Planck scale.
There are strong prejudices about the form that the four dimensional
low energy effective field theory derived from superstrings must take in order
 for superstrings to be a viable model of particle interactions.
In this paper, we extract the most generic elements of the ``current
superstring
lore" (CSSL) and examine their implications for the early evolution of the
universe.

Our goal here is to be strict and systematic in bringing
 CSSL and cosmology together. Hence,
it is necessary  to  carefully  lay down the components of
  CSSL and to not deviate from them.
In the past, several authors have encountered some of the cosmological
problems of superstrings outlined below and have resorted to adding
new terms in the effective potential to solve them \cite{bg}-\cite{vt}.
However,  these added terms
violate one or more key tenets of present superstrings lore;
the cosmology problems of CSSL are thereby masked.
The main components of  CSSL  are:

\noindent
(1)~The effective theory
below the Planck scale is described by a weakly-coupled,
four-dimensional,  $N=1$ supergravity  field theory.
Hence, the effective action is
Kahler invariant, described by a real Kahler potential, $K(\phi_i)$,
and a holomorphic
superpotential $W(\phi_i)$:
\begin{equation}
{\cal L}_{\rm eff} =\sum_{i,j} {1\over 2} K_{\phi_i,{\bar\phi}_j}
\nabla \phi_i\cdot \nabla{\bar\phi}_j
 - {\rm e}^K \left[\  \sum_{i,j } K^{\phi_i\ {\bar\phi}_j}
(D_{\phi_i} W)(D_{{\bar\phi}_j} W)^{\dag} - 3 |W|^2\ \right]
 \end{equation}
where the sum is over (complex-valued) chiral superfields
fields $\phi_{i}$, and
 $D_{{\phi_i}} =\partial_{\phi_i}-K_{\phi_{i}} $,
 $K_{\phi_i} = \partial_{\phi_i} K$ and
$ K^{\phi_i\ {\bar\phi}_j}= K^{-1}_{\phi_i\ {\bar\phi}_j}$.
The Kahler potential,
\begin{equation}
K = - {\rm ln} \; (S + S^*)  \; -3 {\rm ln} \; (T+T^*) + \ldots
\end{equation}
includes two fields:
$S$ (related to the dilaton) and $T$ (related to
the breathing mode). According to CSSL, the contributions of other moduli
(fields similar to $T$), matter fields and higher order corrections
not shown explicitly here can be expressed as corrections to $K$ and $W$.

\noindent
(2)~ The  unified gauge coupling constant is fixed at
 small value $g_{GUT}\approx .7$ to agree with recent hints from LEP
suggesting that, within the  minimal supersymmetric standard model,
 a unification of coupling constants occurs  at a scale
$M_{GUT}\approx 10^{16}$ GeV. The gauge coupling constant is determined
by the expectation value of the real part of $S$, $<Re S>\sim {1\over g^2}>1$.

\noindent
(3)~The source of dilaton interactions is in the hidden sector of
the theory and appears at some intermediate scale $\Lambda_c\approx
10^{14}$~GeV.
The dilaton  and moduli interactions induce
supersymmetry breaking in the observed sector of the theory
\cite{dixon}. Supersymmetry  breaking in the observed sector occurs at
approximately the electroweak scale  $10^3$~GeV.
The hierarchy between the scale $\Lambda_c$ of the hidden sector
 and supersymmetry breaking in the observed sector,  is created because
of the weakness  of the induced dilaton interactions.

\noindent
(4)~The dilaton potential is set by non-perturbative interactions.
An important  consequence of Condition~(3) above is that the
dilaton potential vanishes at temperatures between the Planck
scale ($10^{19}$~GeV) and the intermediate scale, $\Lambda_c \approx
10^{14}$~GeV.
Below $\Lambda_c$, the superpotential is of the form:
\begin{equation}\label{1}
W(S,T) = \sum_{j} {\rm e}^{-\alpha_{j} S} f_{j}(S,T,\ldots),
\end{equation}
where  $\alpha_{j}$ are constants and $f_j$  depend on matter fields,
and  may contain powers of $S$.
The potential is,
using Eq.~(1):
\begin{equation} \label{potent}
V(S)=(S + S^{*}) | \partial_S W(S) - \frac{1}{S+S^{*}}W(S)|^{2}-
\frac{3}{S+S^{*}}|W(S)|^2
\end{equation}
 The dependence on $T$  and other moduli is not shown explicitly here
because their expectation values are fixed  by modular invariance
\cite{mc}.

Although  some of our arguments depend only on the general, non-perturbative
form of the superpotential, for concreteness
 we will assume the current leading candidate as a source of
this non-perturbative potential: gaugino condensation \cite{ggino},\cite{kras}.
Above the scale of gaugino condensation, $\Lambda_c$, the superpotential
is identically zero.  Below $\Lambda_{c}$, the superpotential has the
form given in Eq.~(\ref{1}).  In this case,
$\alpha_{j} = 3 k /2 \beta$, where $k$ is a positive integer,
$\beta$ is determined by the one-loop
beta-function of the hidden gauge group, and the $f_j$'s are  known constants,
determined by the structure of the hidden gauge group and the potential for
the $T$ field and matter fields \cite{ccm}.
\beq\label{2}
W(S) = \sum_{j} a_j {\rm e}^{-\alpha_{j} S}
\eeq
If the gauge group is simple there is only one term in the sum.
If the gauge group is  semi-simple, each additional term corresponds to
gaugino condensation in  one of  the distinct factors of the group.
The  rank of the gauge group is restricted to be at most 18.
The sum in Eq.~(\ref{2}) is therefore a small finite sum.

Now we wish to show how this prevalent view of superstrings runs into
immediate problems  once cosmology is considered.  First, we shall
 argue how  dilaton interactions create a serious roadblock
 for inflationary cosmology.  Since inflation is not yet a proven
 idea, this problem is arguably the least severe.  Then, we shall
 see how the  anticipated dilaton interaction
 makes it highly unlikely that the universe ever reaches a state
 with zero (or small) cosmological constant and a state in which
 Newton's constant $G$ is time-independent, $\dot{G}=0$.  Here, the
contradictions are with strong observational limits; e.g., see
  \cite{ccp}-\cite{cgbq}.
In the end, we  do not claim
 a rigorous theorem.  However, we have been unable to identify
 a viable model,  and  we seem
 to be driven to unattractive extremes (at best).

  Any type of inflationary scenario requires an epoch in which the
  false vacuum (potential) energy density dominates the energy
  density of the universe \cite{infsc}. The  false vacuum energy
 density   must drive the scale factor $a(t)$  to grow  as
   $a(t) \propto t^{m}
  $ where $m\ga 4$ \cite{eia} in order to solve the cosmological
  flatness, horizon, and monopole problems.
The anticipated dilaton interaction can prevent the universe from
undergoing superluminal expansion in two ways.
Before the dilaton settles into a stable minimum of its potential,
the dilaton kinetic energy density  can dominate the vacuum energy
density.
 Once the
dilaton settles at a minimum and its kinetic energy becomes negligible,
the problem  may be that there is not sufficiently
 large potential energy density
to drive inflation.

Let us first consider the problem before the dilaton
settles to a minimum.  Numerical integration is required to obtain
quantitative results. However, there is a useful analogy to Brans-Dicke
theory that allows us to explain the essence of our results in a simple way.
Both a Brans-Dicke scalar and  the dilaton are
non-minimally coupled to the scalar curvature.
 Through its non-minimal coupling, the
 Brans-Dicke field $\phi$  is subject to
a force proportional to the false vacuum energy density.
 Instead
of the vacuum energy density being focused totally into inflating
the scale factor $a(t)$,
 some of it is funneled off into driving the $\phi$ field \cite{eib}.
The kinetic energy density of the $\phi$ field changes the overall equation of
state,  slowing down the expansion
rate (i.e., reducing $m$).

If the Brans-Dicke field is Weyl transformed so that the scalar
curvature term assumes standard Einstein form ($(16 \pi M_{Pl}^2)^{-1}
{\cal R}$),
 $\phi$ has
canonical kinetic energy density, but the false vacuuum energy
density $V_F$ is modified by $\phi$-dependent factor,
$V(\phi) = V_F \; {\exp} (-\beta \phi/M_{Pl} )$,
where $\beta =
\sqrt{64 \pi/(2 \omega +3)}$ and  $\omega$ is the Brans-Dicke
parameter that determines the deviation from Einstein gravity.
The equations of motion for the Weyl-transformed theory are:
\begin{equation}
\ddot{\phi} + 3 H \dot{\phi} = - \frac{d V(\phi)}{d \phi},
\end{equation}
where
$H^2 \equiv ( \dot{a}/a)^2 =
 8 \pi (\frac{1}{2} \dot{\phi}^2 + V)/(3 M_{Pl}^2)$.
 The solution is $a(t) \propto t^{(2\omega+3)/4}$
 so that  $\omega\ge 5$ is required to resolve the
 horizon and flatness problems. A slightly weaker bound $\omega > 1/2 $
is required to have  any superluminal expansion \cite{eib}.
For  $\omega<1/2$,
 the potential for $\phi$ in the Weyl transformed theory is so steep
 that the $\phi$ kinetic energy density grows to dominate
 any potential energy density.

In superstrings, the dilaton $\phi$ is related to $S$ by $Re \; S = {\rm
e}^{\phi}$.
The effective potential is identically
zero until some intermediate scale
$\Lambda_c\approx 10^{14}$~GeV.  Since inflation requires a non-zero vacuum
energy
density, inflation can only occur after a non-zero potential is generated
below $\Lambda_c$. Below $\Lambda_c$, the dilaton potential assumes the form
 given in Eq.~(5). Any prospective inflaton must couple to $S$. Under the
assumption that the dilaton potential is strictly non-perturbative, the false
vacuum energy must depend on $S$ through one or more terms in Eq.~(5). Hence,
the following argument is independent of the type of inflationary model (e.g.,
chaotic, extended, etc.).

In  Figure~1, we plot some typical dilaton potentials corresponding to
one and two gaugino condensates.
Note that the potential is unbounded below for $\phi <0$; this corresponds
to the   strong coupling region, where the non-perturbative
 analysis that led to this potential cannot be trusted.  Hence, we
 will restrict ourselves to  $\phi>0$.  The striking feature of  the
 potential is its steepness.  Because  the  potential is non-perturbative
 in origin, the potential drops as ${\rm e}^{-2 \alpha S} \sim {\rm
e}^{-2\alpha
 {\rm e}^{\phi}}$,
 for large $\phi$, where $\alpha$ is one of the coefficients in Eq.~(\ref{2}).

 \vfill\eject
%\centerline{\psfig{file=dilpictI.epsf.ps,height=3.8in,width=4.9in}}
%\epsfscale=900 \centerline{\ \ \ \epsfbox{dilpictI.epsf}}
{\ \ \ \  }
\vspace{3.2in}

{\small Figure 1. Non Perturbative Dilaton Potential $V(\phi)/ M_{Pl}^4$.
The dashed line   \hfill\break\hbox to 42pt{\hfil} corresponds to one
gaugino condensate and the solid line corresponds
\hfill\break\hbox to 42pt{\hfil} to two gaugino condensates. The insert shows
 $V(\phi)$  in the region where
\hfill\break\hbox to 42pt{\hfil} the potential resulting from two gaugino
condensates has a minimum,\hfill\break\hbox to 42pt{\hfil}
 scaled to  show the minimum and the steepness of the potential.}

\noindent
 The  precise value of $\alpha$ depends on the hidden gauge group, but  among
 the known examples we find $\alpha \ge 24 \pi^2/r$, where $r$ is
 an integer number of order ${\cal O}(10)$.
 Expanded about any $\phi_0>0$ ($S_0 \equiv {\rm e}^{\phi_0}>1$),
 the potential can be approximated by
 ${\rm exp}(-2\alpha S_0 (\phi-\phi_0))$. Comparing this with the Brans-Dicke
 theory, we find a small effective
$\omega \approx 4.5\times 10^{-4} (r/S_0)^2 -1.5$.
 For typical hidden sector groups,
   $\omega <1/2$, a value too small for inflation owing to the steepness
   of the potential; for $E_8$, we find the maximal value $r=90$ which,
   combined with $S_0 =1$ squeaks past superluminal expansion with
   $\omega<2$, but still $\omega$ is too small to solve the cosmological
    horizon and flatness problems.
 Consequently, {\it until $\phi$ settles near a minimum of
 its potential, the dilaton kinetic energy  totally dominates the
 potential energy density, thereby blocking inflation of any
 kind}.\footnote{Strictly speaking, there can  be inflation if $\phi$
happens to begin  very close to the peak of one of the small hills,
e.g., see  the two-gaugino potential in Fig.~1 . However,
this  seems to be an extraordinarly fine-tuned and unattractive
initial condition.}

The best-hope for inflation then becomes a scenario in which the
dilaton first rolls to a minimum, and then inflation is driven by
other  fields.
There are two problems with this approach, though.
First, as is apparent from Fig.~1, for $\phi$
be trapped in any minimum some very strict constraints on the initial
conditions
of $\phi$ have to be satisfied. Initially (above $\Lambda_c$) there is no
 potential and it seems reasonable to assume that $\phi$ is  has an arbitrary
value that is spatially varying. If $\phi$  begins to the right of
the ``bumps'' in the potential, it  rolls continuously towards
$\phi \rightarrow \infty$.
If $\phi$ begins close to $\phi=0$, the potential is so
steep that $\phi$ rolls right over the bumps and continues onto
$\phi\rightarrow \infty$.  Only  for a narrow region about the minimum
will $\phi$ be properly trapped.  However, this condition is not
sufficient: the energy density must be positive to have inflation,
whereas the example in Fig.~1 has  a minimum with negative vacuum
energy density.  The minimum must also be  {\it metastable} since,
after inflation, the universe needs to find its way down to  a lower
energy state with zero cosmological constant.
 Thus far, we have been unable to construct a superpotential of the form in
 Eq.~(\ref{2}) which has the desired properties.

Even without inflation, we have the embarrassing problem of understanding
why $\phi$ gets trapped  at a minimum rather than rolling past the
bumps and evolving forever.
  The dilaton must be trapped
because, otherwise, its evolution leads to an unacceptably large
time-variation of Newton's constant $G$ \cite{ds},\cite{cgbq}.
As we have argued, so long as $\phi$ is
unpinned, the effective $\omega$ is less than unity,
whereas cosmology demands $\omega$
to be greater than  50 in order to  meet the constraints of
primordial nucleosynthesis and solar system tests (e.g., Viking
radar-ranging) push the constraints even higher, $\omega>500$ \cite{will}.
At this point, the only means of trapping $\phi$ is
if its  initial value is close to one of the minima in the first
place. It would be nice if some natural damping mechanism existed which forces
the
dilaton to settle in its shallow minimum.

But is the minimum of a dilaton potential
where  the  universe really ought to be?  The problem
here is that it is very difficult\footnote{
A special mechanism for obtaining
zero cosmological constant \cite{ccp},\cite{bq} has been used in the
literature \cite{ggino},\cite{bg} in which a field
appears in the Kahler potential but not in the
superpotential. This approach
necessarily leads to a flat direction in the
potential and a massless field. It is unknown if there
is a mechanism to give the field a mass and yet
maintain zero cosmological constant.}
avoid a  minimum with  negative cosmological constant!  This
was already a problem in generic supergravity theories \cite{bp};
 the problem is  exacerbated  when one considers the more
 special non-perturbative
 potential  in Eq.~(\ref{2}).

A minimum of the potential satisfies $\partial_S V(S)=0$
and can be supersymmetric in the $S$ sector,
$ D_S W = 0$, or produce  supersymmetry breaking, $ D_S W \ne 0$, where $D$ is
the Kahler derivative defined below  Eq.~(\ref{1}).
It is straightforward to show that any extremum  of $V(S)$
 in the weak coupling region (namely, $ReS>1$, where the desirable minimum
should be according to condition 2 of the CSSL), with $W(S)$  of the form in
Eq.~(\ref{2}) and $ D_S W \ne 0$ is a saddle point of $V(S)$, rather than
a stable minimum \cite{ccm}.  Hence,  the present universe could not be in a
vacuum state which is  supersymmetry breaking in the $S$ sector.

The only remaining possibility is that the present state of the universe
corresponds to a minimum with $ D_S W = 0$:
either
$\partial_S W(S)|_{S_{min}} = W(S_{min})=0$ (a minimum of
Type~A);  or $W(S_{min})$ and
$\partial_S W(S)|_{S_{min}}$ are both non-zero (a minimum of
Type~B). If supersymmetry were unbroken in the
 whole  theory then the potential in Eq.(\ref{potent}) is exactly correct
 and  because $ D_S W = 0$ but $W(S_{min} )\ne 0$ in  Type~B, $V(S_{min})$
is negative (see Eq.~(\ref{potent})) and Type~B minima would necessarily
have an unacceptable negative cosmological constant.  However,
supersymmetry has to be broken in some sector the theory.
 This means that the factor 3, in Eq.(\ref{potent}) gets modified to a
smaller number $n<3$. In  all presently known
examples,
supersymmetry breaking occurs in the moduli sector \cite{mc},\cite{ccm}
 and $n>1$ always.  Provided $n>1$, the conclusion that  the potential
is negative at Type~B minima is not modified.

The last hope
for a realistic vacuum state is a  minimum of Type~A.
However, we shall now argue that even if one is able to arrange a minimum
with zero cosmological constant of Type~A, it is almost always accompanied
by a lower energy minimum of Type~B with negative cosmological constant in
which the
universe is more likely to be trapped. This is the situation
depicted in Figure 2, it corresponds to an example found in ref.\cite{taylor}.
We ignore here contributions from supersymmetry breaking in other sectors of
the theory.

%\centerline{\psfig{file=dilpictIII.epsf.ps,height=2in,width=2.5in}}
%\epsfscale=600 \centerline{\ \ \ \epsfbox{dilpictIII.epsf}}
\vspace{2.0in}

{\small Figure 2. Dilaton potential corresponding
 to gaugino condensation in  \hfill\break\hbox to 42pt{\hfil} the hidden
 group  $G=SU(2)_{k=1}\times \biggl[SU(2)_{k=2}\biggr]^4
\times\biggr[SU(2)_{k=3}\biggl]^4$.}

\noindent
 First, let us consider
the case where the coefficients in Eq.~(\ref{2}) are all real and there
is a minimum of Type~A for some real value of $S=S_A$.
For any hidden gauge group, the coefficients $\alpha_j$ in Eq.~(\ref{2})
can all be written as $n_j \tilde{\alpha}$, where  the $n_j$'s are integers.
Hence, if we set
 $z={\rm  e}^{-{\tilde\alpha} S}$, $W(S)$ becomes a polynomial $P(z)$.
 As $S \rightarrow \infty$, $W(S) \rightarrow 0$  exponentially
 for superpotentials of
 non-perturbative form; consequently, $P(z=0) =0$. If $z_A ={\rm e}
 ^{-\tilde{\alpha}  S_A}$ (which is real by our assumptions),
 then, since $W(S_A)=\partial_S W(S_A)=0$, we
 can write
 $P(z)=z(z-z_A)^2 \widetilde P(z)$. There may be many Type~A minima along
 the real axis, but
 let us assume that $S_A$ is the most positive value.
Now let us consider $P(z)$  between $z=z_A$ and $z=0$ (corresponding to
$S_A < S<\infty$).
 Rolle's theorem says that between every two zeros
of a real function lies a zero of its derivative. Since $P(z)$ has a zeros
at $z=z_A$ and $z=0$, this means that whenever we
have a minimum of Type~A, there will be a
 point between, call it $z'$, corresponding to $S'>S_A$,
 where $P'(z')= 0$ or $\partial_S(W) =0$.
If we also had $W=0$ and this were a minimum, then it would be of Type~A,
contradicting our assumption that $S_A$  is the most positive
minimum of Type~A. That leaves two possibilities: (i)~$W\ne 0$, in which
case $V(S')<0$ (see Eq.~(\ref{potent}); or (ii)~$W=0$,  and, hence, $V(S')=0$,
but  $S'$ is not a local minimum.  In either case, we see that
{\it there must be a range of the effective potential with negative
$V(S)$, in the neighborhood of $S'$.} This means that the global minimum of
$V(S)$ is negative. In practice, these are  relatively deep minima which the
universe can avoid only for special initial conditions.

If we consider complex coefficients in Eq.~(2),
there is a useful, albeit somewhat weaker generalization
of Rolle's theorem that applies  specifically to polynomial
$P(z)$ \cite{marden}.   We begin with the same assumptions that $P(z)$ has
a  zero at $z=0$ ($S\rightarrow \infty$) with multiplicity $k_0\ge 1$ and
a multiplicity $k_A\ge 2$ zero (Type~A minimum) at some  real $z=z_A$.
Then, there is at least one additional $z=z'$ in the complex
plane for which $P'(z')=0$.  The point $z'$
lies within a bounded  region ${\cal A}$ about the segment ${\overline {OA}}$
 which joins $z_0$ and $z_A$.  The same reasoning as above implies
 that  there is a neighborhood of $z'$ where  $V(S)<0$, a potentially
 dangerous region of the potential with negative cosmological constant.
 Region ${\cal A}$ for polynomial $P(z)$ of order $n$
 is the union of  the two circles for which ${\overline {OA}}$
is a chord subtending  an angle $2 \pi/(n+1-k_0-k_A)$.  For
fixed $k_0$ and $k_A$, ${\cal A}$ grows as $n$ increases.  For
 concreteness, let us consider the minimal case, $k_0=1$ and $k_A=2$.
 Then, for $n\le4$, ${\cal A}$ lies totally in  the physically allowable
 region ${\rm Re }\; S
 >0$ ($|z|<1$); in this case,
  the region of $V(S)$ with negative cosmological constant
  near $z'$ is a real danger.   However, for sufficiently large
  $n>4$ (depending on the value of $z_A$), ${\cal A}$ extends to
 the unphysical region ${\rm Re}\; S <0$ ($|z|>1$).  Hence,
 for sufficiently large $n$, it is conceivable that $V(S)$ only
 has Type~A minima in the physical region. So, the situation is
 not completely hopeless, but one must pay a heavy price:
 (a)~the coefficients $a_j$ in Eq.~(\ref{2}) must be complex, or
equivalently, $<ImS>\ne 0$, which may not be possible  for realistic
 CP invariant theories;
  and (b)~in order to satisfy $n>5$  for $k_0=1$ and $k_A=2$
  (or, more generally, $n>2+k_0+k_A$) so that the Type~B minima
  are pushed into the unphysical region, at least five gaugino
  condensates are required.  This combination of conditions seems
  unwieldy, unlikely and unattractive.

We are, therefore, forced to conclude  that
it seems  difficult to construct
a model  consistent with current superstring lore which can support
inflation or which can lead to  a vacuum state with zero cosmological constant
and time-invariant Newton's constant.
While we have pointed out some exotic loopholes entailing special
initial conditions and complex superpotentials,
we would recommend against this
unattractive approach. Rather, we have presented this discussion
because we believe that  superstring
theorists should be aware of  the impending
cosmological disasters and take up our challenge:
 Is there a plausible  modification
of superstring lore that will  fit better with cosmology, and does this
modification give us a promising implications for  particle physics?

{ACKNOWLEDGEMENT}:
It is a pleasure to thank  Charles Epstein, Jens Erler, Vadim Kaplunovsky,
 Burt Ovrut and Steven Weinberg for useful discussions and Mirjam Cvetic for
collaboration in the early stages of this work.
This work was supported in part by the Department of Energy under
contract No. DOE-AC02-76-ERO-3071.
\end{document}